\documentclass[useAMS,usenatbib]{mn2e}
\usepackage{graphicx} \usepackage{rotating}
\usepackage{txfonts}
\usepackage{url} 
\usepackage{fmtcount}
\usepackage{upgreek}

\def\nt/f{Nuclear Technology/Fusion}

\title[$^3$He: Does the problem persists$?$]
{\begin{center}
$^3$He: Does the problem persists$?$
\end{center}}
\author[L. Guzman-Ramirez et al.]{L. Guzman-Ramirez$^{1}$\thanks{E-mail: lizette.ramirez@postgrad.manchester.ac.uk}, J.~E. Pineda$^{1}$, A.~A. Zijlstra$^{1}$, R. Stancliffe $^{2}$ and A. Karakas$^{3}$\\
$^{1}$Jodrell Bank Centre for Astrophysics, School of Physics and Astronomy, University of Manchester, Manchester, M13 9PL, UK\\
$^{2}$Argelander-Institut f\"ur Astronomie, Auf dem H\"ugel 71, D-53121 Bonn, Germany\\
$^{3}$Research School of Astronomy \& Astrophysics, Mt Stromlo Observatory, Canberra, Australia
}

\date{Released 2013 Xxxxx XX}

\pagerange{\pageref{firstpage}--\pageref{lastpage}} \pubyear{2002}

\def\LaTeX{L\kern-.36em\raise.3ex\hbox{a}\kern-.15em
    T\kern-.1667em\lower.7ex\hbox{E}\kern-.125emX}

\begin{document}

\date{Accepted 2013 March 20. Received 2013 March 14; in original form 2013 February 25}

\pagerange{\pageref{firstpage}--\pageref{lastpage}} \pubyear{2013}

\maketitle

\label{firstpage}

\begin{abstract}
To understand the chemical evolution of the Galaxy, we need to understand the contribution of PNe in the $^3$He abundance. $^3$He abundances have been detected only in a couple of PNe, their abundances are consistent with the standard models, in which $^3$He  is produced in significant quantities by stars of 1--2M$_{\odot}$. However, for all the other PNe observed to date there have been no detections, therefore only upper limits in their abundance can be calculated. 
Observations of the $^3$He$^+$ emission line using the {\it VLA} towards the PNe IC 418, NGC 6572 and NGC 7009 were used to obtain upper limits for their $^3$He abundance. 
Because the abundance of $^3$He in H\,{\sc ii} regions, the ISM and the proto-solar system is much lower than what  is predicted, new chemical models were proposed. The resulting evolution of $^3$He, based on stellar evolution models, can be consistent with the values determined in pre-solar material and the ISM if 96\% of the population of stars with mass below $\sim$2.5M$_{\odot}$ has undergone enhanced $^3$He depletion. This implies that unless the combined sample of PNe that has been observed so far is very atypical, the solution to the ``The $^3$He Problem" lies elsewhere. However, the results presented here suggest that more observations are needed in order to make a strong conclusion about the stellar evolution models. 
\end{abstract}

\begin{keywords}
circumstellar matter -- radio: abundances, planetary nebulae.
\end{keywords}

\section{Introduction}
The cosmic abundance of the $^3$He isotope has important implications for many fields of astrophysics. It impacts stellar evolution, chemical evolution and cosmology. The present interstellar $^3$He abundance, as per all the light elements, comes from a combination of Big Bang Nucleosynthesis (BBN) and stellar nucleosynthesis \citep{wilson94}. H\,{\sc ii} regions are young objects compared with the age of the Galaxy, so they represent the zero-age objects. Their $^3$He abundance is the result of 13.7Gyr of Galactic chemical evolution. Any hydrogen-burning zone of a star which is not sufficiently hot ($>$7$\times$10$^6$K) will produce $^3$He via the {\it p-p} chain. Stars with masses $<$2.5M$_{\odot}$ are the net producers of $^3$He. For these stars, {\it p-p} burning is rapid enough to produce D {\it in situ}, and enable the production of $^3$He (D+ p $\rightarrow$ $^3$He + $\gamma$). More massive stars are dominated by CNO burning; although they do produce some $^3$He via the {\it p-p} chain in their outer zones, it is not enough to offset the $^3$He destruction in the interior zones \citep{dearborn86}. Therefore PNe evolved from stars with masses $<$2.5M$_{\odot}$ should be substantially enriched in  $^3$He. PNe  $^3$He abundances are important tests of stellar evolution theory \citep{bania10}. 

\citet{galli95} presented  ``The $^3$He Problem". According to standard models of stellar nucleosynthesis there should be a $^3$He/H abundance gradient in the Galactic Disk with the highest abundances occurring in the highly-processed inner Galaxy,  the $^3$He/H abundance should grow with source metallicity and the protosolar $^3$He/H value should be less than what is found in the present ISM.

After more than 30 years of effort, observations of  the abundance of $^3$He in H\,{\sc ii} regions are inconsistent with these expectations \citep{rood79,bania07}. For the $^3$He problem to be solved, it will require that the vast majority of low-mass stars fail to enrich the ISM. One suggestion to solve this is adding extra mixing in the red giant branch (RGB) stage. This extra mixing adds to the standard first dredge-up to modify the surface abundances. \citet{eggleton06} showed that mixing in the surface convective zone of low-mass stars in the RGB phase is extended below the classical convective limit and that it is very fast compared with the nuclear time scales of either the hydrogen-burning shell or the $^3$He-burning shell. Their model estimated that as the H-shell burns outwards the $^3$He will be destroyed. Although low-mass stars do produce considerable amounts of $^3$He on the main sequence phase, this will be all destroyed by the substantially deeper mixing they expect to occur in the red giant branch phase. These same results estimate that while 90\% of the $^3$He is destroyed in 1M$_{\odot}$ stars, only 40-60\% is destroyed in a 2M$_{\odot}$ star model, depending on the speed of mixing. For stars above 4M$_{\odot}$, the bottom of the convective envelope is hot enough for nuclear burning (hot bottom burning). This leaves a mass range around 2M$_{\odot}$ which is still expected to contribute to $^3$He enrichment. Recent observations of \citet{smiljanic09} have further complicated the issue: they found a few 2.5--4M$_{\odot}$ stars with low $^{12}$C/$^{13}$C ratios, where CN-processed material has been brought from the H-shell to the envelope. Extra mixing mechanisms do not appear to operate in these stars, and additional mechanisms are assumed. Whether these would affect $^3$He is not known.

The abundance of $^3$He can only be derived from measurements of the hyperfine transition of $^3$He$^+$ which has a rest frequency of 8.665 GHz. Detecting $^3$He$^+$ in PNe challenges the sensitivity limits of all existing radio telescopes. The detection of $^3$He in any PN will require $^3$He/H$\sim$10$^{-4}$, which is the abundance predicted by standard stellar models. After an intense search of $^3$He in PNe, the results are still unclear. \citet{bania10} observed a sample of 12 PNe. $^3$He was detected in only 2 of them. NGC 3242 was observed with Effelsberg a 100m dish from the Max Planck Institute for Radio Astronomy (MPIfR) and the NRAO 140-foot telescopes, however the observations are inconsistent with each other, implying a $^3$He/H abundance that is $\sim$25\% of the 10$^{-4}$ abundance predicted. For the case of J320, $^3$He was detected at a 4$\sigma$ level with the NRAO {\it VLA}. Composite $^3$He$^+$ average spectrum for 6 PNe (NGC 3242, NGC 6543, NGC 6720, NGC 7009, NGC 7662, and IC 289), using Effelsberg, Arecibo and {\it GBT} observations, consistently show $^3$He$^+$ emission at the $\sim$1mK level.

$^3$He is created in low mass stars and ejected during the late phases of their evolution. Galactic models predict a rapid increase over the past 5Gyr. This is however in contradiction to the observed abundances in the ISM. The cause is believed to be destruction of $^3$He during the RGB phase, by hot bottom burning at higher masses and thermohaline mixing at low masses, leaving only a narrow mass range between. In this paper we present upper limits estimates for 3 PNe. The description of the observations made using the {\it VLA} in Section 2, the abundance calculations and results are presented in Section 3, the stellar models are presented in the discussion in Section 4 and the main conclusions are in Section 5.

\section{Observations using the {\it VLA}}

Motivated by this problem, three planetary nebulae were observed using the national radioastronomy observatory (NRAO) very large array ({\it VLA}) in D configuration. In Table \ref{obsPara} the targets and the parameters of the observations are presented.

\begin{table*}
\centering
\caption{{\it VLA} Observations Parameters.}
\label{obsPara}
\begin{tabular}{lcccc} 
\hline\hline Parameters			& 	\,	&	IC418	&	NGC 6572	&	NGC 7009 \\
\hline
Date$^{\star}$					&	\,	&15-12-2009	&	13-12-2009	&26-12-2009 \\
Total observed time (hr)			&	\,	&	6		&	6.2			&	4.6		\\
Configuration					&	\,	&	D		&	D			&	D		\\
R.A. (J2000)					&	\,	& 05:27:28.20	& 18:12:06.36		&	21:04:10.8 \\
DEC (J2000)					&	\,	& --12:41:50.26	&  +06:51:13.01	&	 --11:21:48.25\\
FWHM of primary beam (arcmin)	&	\,	&	5.4		&	5.4			&	5.4		\\
Size of the PNe (arcsecs)			&	\,	& 30$\times$6	& 18.9$\times$6.9	&  12$\times$7	\\
FWHM of synthesised beam (arcsec)&	\,	&	13		&	9			&	9		\\
Total bandwidth (MHz)			&	\,	&	6.152	&	6.152		&	6.152	\\
Number of channels				&	\,	&	62		&	62			&	62		\\
LSR central velocity (km\,s$^{-1}$)	&	\,	&	62.0		&	--8.7			&	--46.6	\\
Observed $\nu$ (MHz)			&	\,	&8662.6338	&8665.6132		&8666.0602	\\
Spectral resolution (km\,s$^{-1}$)	&	\,	&	3.378	&	3.378		&	3.378	\\
Flux density calibrator			&	\,	& 0137+331	& 1331+305		& 1733--130	\\
Phase calibrator				&	\,	& 0609--157	& 1751+096		& 2136+006	\\
Line channel rms (mJy\,beam$^{-1}$)$^{\star\star}$	&	\,	&	1		&	0.4			&	0.8		\\
Continuum rms (mJy\,beam$^{-1}$)	&	\,	&	2.0		&	1.53			&	1.3		\\
 \hline
\multicolumn{5}{|l|}{$^{\star}$More observations were obtained, due to the lack of calibrators or bad data they were not considered here.} 
   \\
$^{\star\star}$Continuum subtracted
   \end{tabular}
\end{table*}

The data was edited and reduced using the Common Astronomy Software Applications (CASA) package. The calibration technique was slightly different for every object due to different S/N ratio in the phase and bandpass calibrators.

\begin{itemize}
\item {\bf IC 418.} For this source the phase calibrator (B0609-157) was used as a bandpass calibrator. The integration time was averaged every hour to get the bandpass solutions.  The calibrator B0137+331 was used to calibrate the intensity scale.  The continuum intensity of this source was sufficient to perform the standard self-calibration technique. The self-calibration algorithm was applied to the continuum after the standard phase and bandpass calibration. 

Three loops of self-calibration were applied, the first 2 were only in phase, using the complete time interval for the first one and then a solutions interval of 300s. After this a self-calibration in amplitude and phase was applied using 150s for the solutions interval. Once this was obtained these solutions were applied to the line data, the continuum was subtracted and the clean image was created.

\item {\bf NGC 6572.} For this source the phase calibrator (B1751+096) was used as a bandpass calibrator; the integration time was averaged every hour to get the bandpass solutions.  The calibrator B1331+305 was used to calibrate the intensity scale.  The continuum intensity of this source was sufficient to perform the standard self-calibration technique. The self-calibration algorithm was applied after the standard phase and bandpass calibration. 
Several loops of self-calibration were applied. The first 4 were only in phase, changing the solutions interval from using the complete integration time up to intervals of 21s. After this a self-calibration in amplitude and phase was applied three times, also changing the intervals of the solutions, starting using 150s.  Ultimately the best result for the self-calibration was obtained using a solution interval of 21s in phase only and a solution interval of 30s for the amplitude and phase. Once this was obtained these solutions were applied to the line data, the continuum was subtracted and the clean image was created.

\item {\bf NGC 7009.} For this source also the phase calibrator (B2136+006) was used as a bandpass calibrator; the integration time was averaged every hour to get the bandpass solutions.  The same calibrator was used to calibrate the intensity scale. For this data set, a flux calibrator was not observed. But as mentioned in Table \ref{obsPara} there were more observations taken for these sources. In the case of NGC 7009 a day before of the observations analysed here a flux calibrator was observed; this was used to calibrate the phase calibrator (same for both observations) and the flux obtained was imposed into the phase calibrator for the data set used here. Therefore the same calibrator was used as a phase, bandpass and amplitude calibrator. For this source a self-calibration technique was applied, but it did not improve the data. Therefore the image and spectrum presented here were made using only the standard calibration. Once this was obtained these solutions were applied to the line data, the continuum was subtracted and the clean image was created.
\end{itemize}

The continuum images are completely limited by the finite dynamic range of the strong continuum emission and thus are not limited by thermal fluctuations. The line data, after subtraction of the continuum, are limited by thermal noise. This explains why the continuum rms values are larger than the line data values (see Table \ref{obsPara}).

\begin{figure*}
\centering
\hbox{\includegraphics[width=8.5cm]{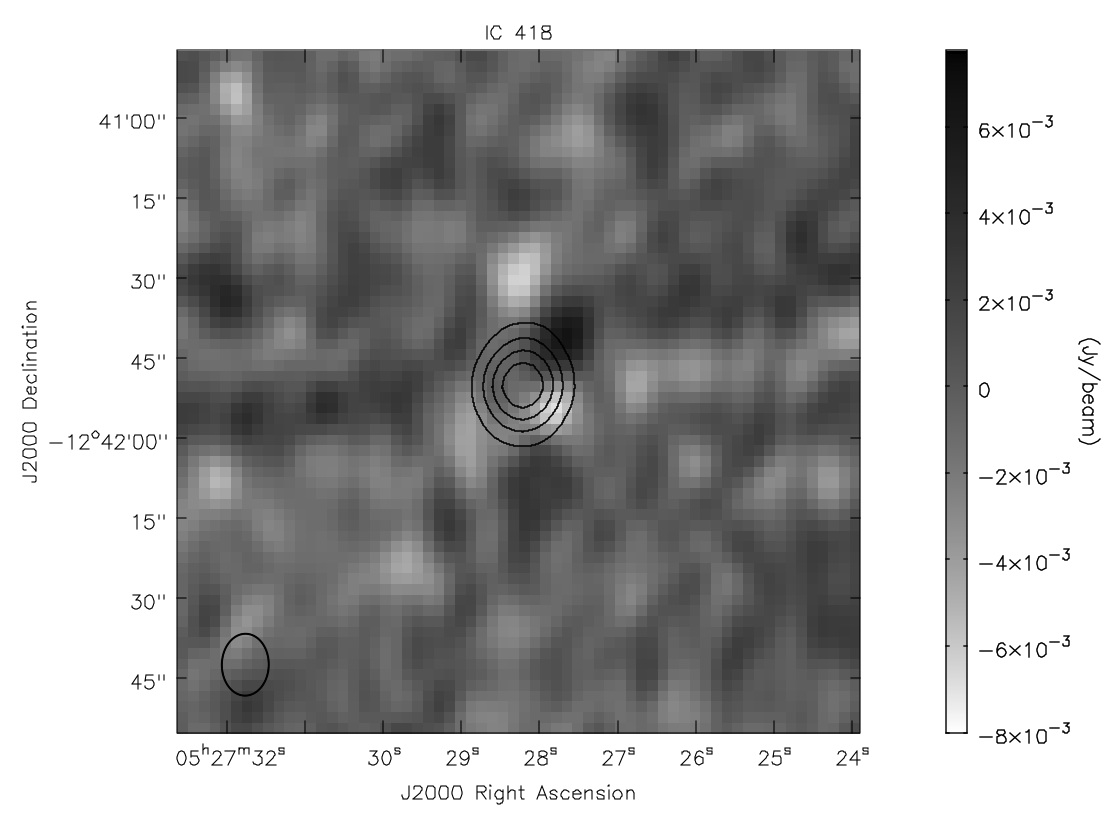}
\hspace{0.1cm}
\includegraphics[width=9cm]{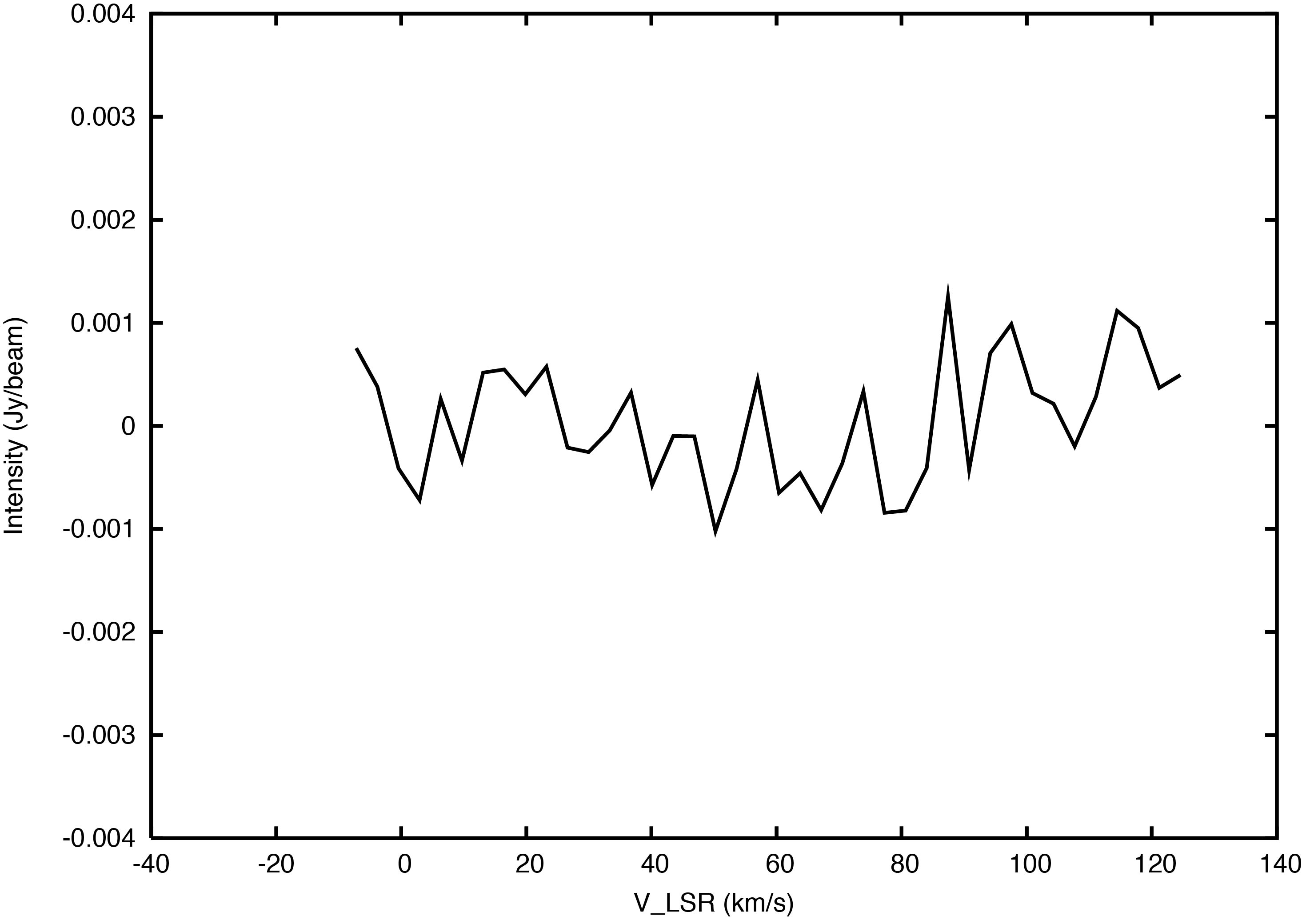}}
\vspace{0.5cm}
\hbox{\includegraphics[width=8.5cm]{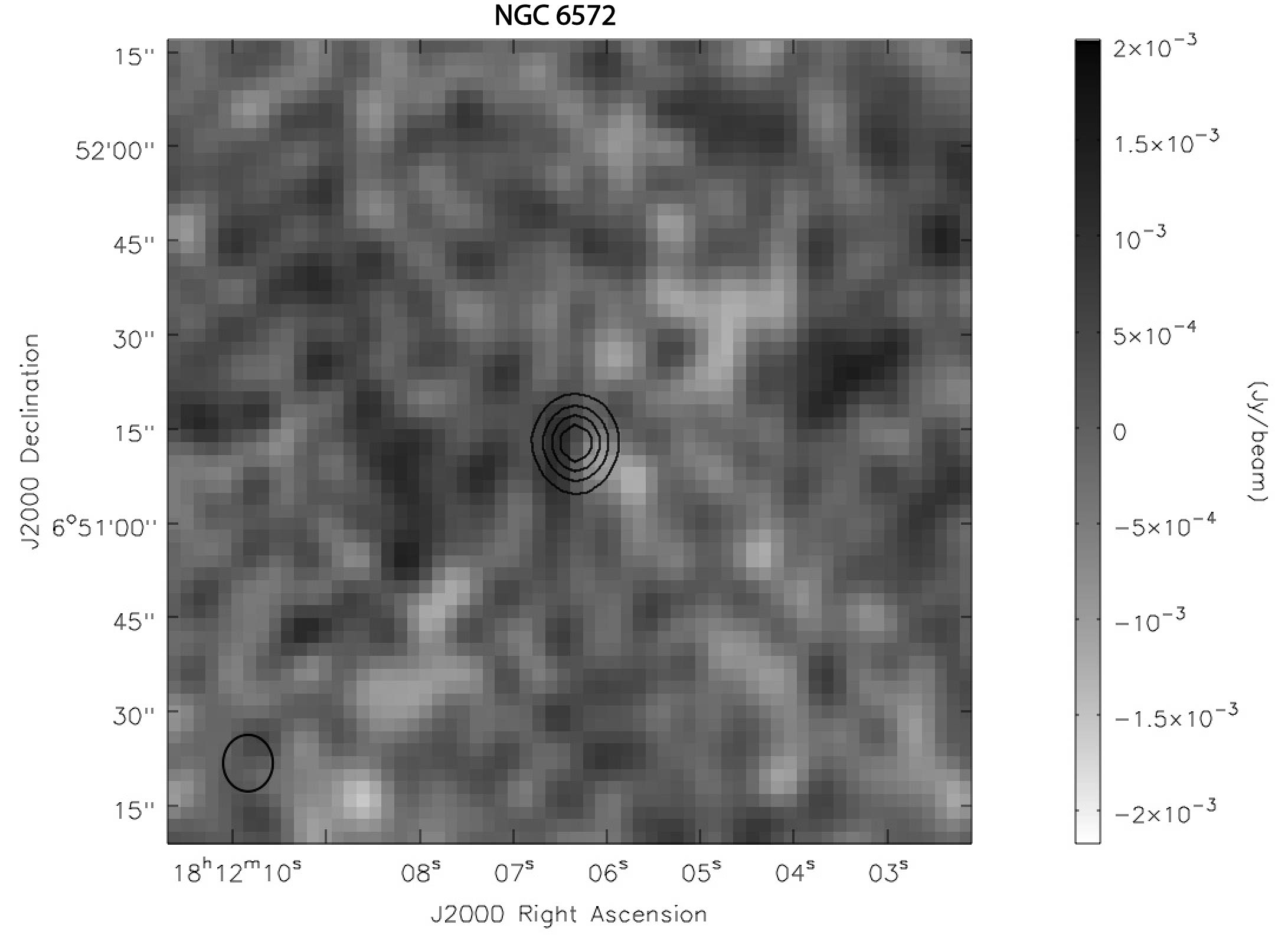}
\hspace{0.1cm}
\includegraphics[width=9cm]{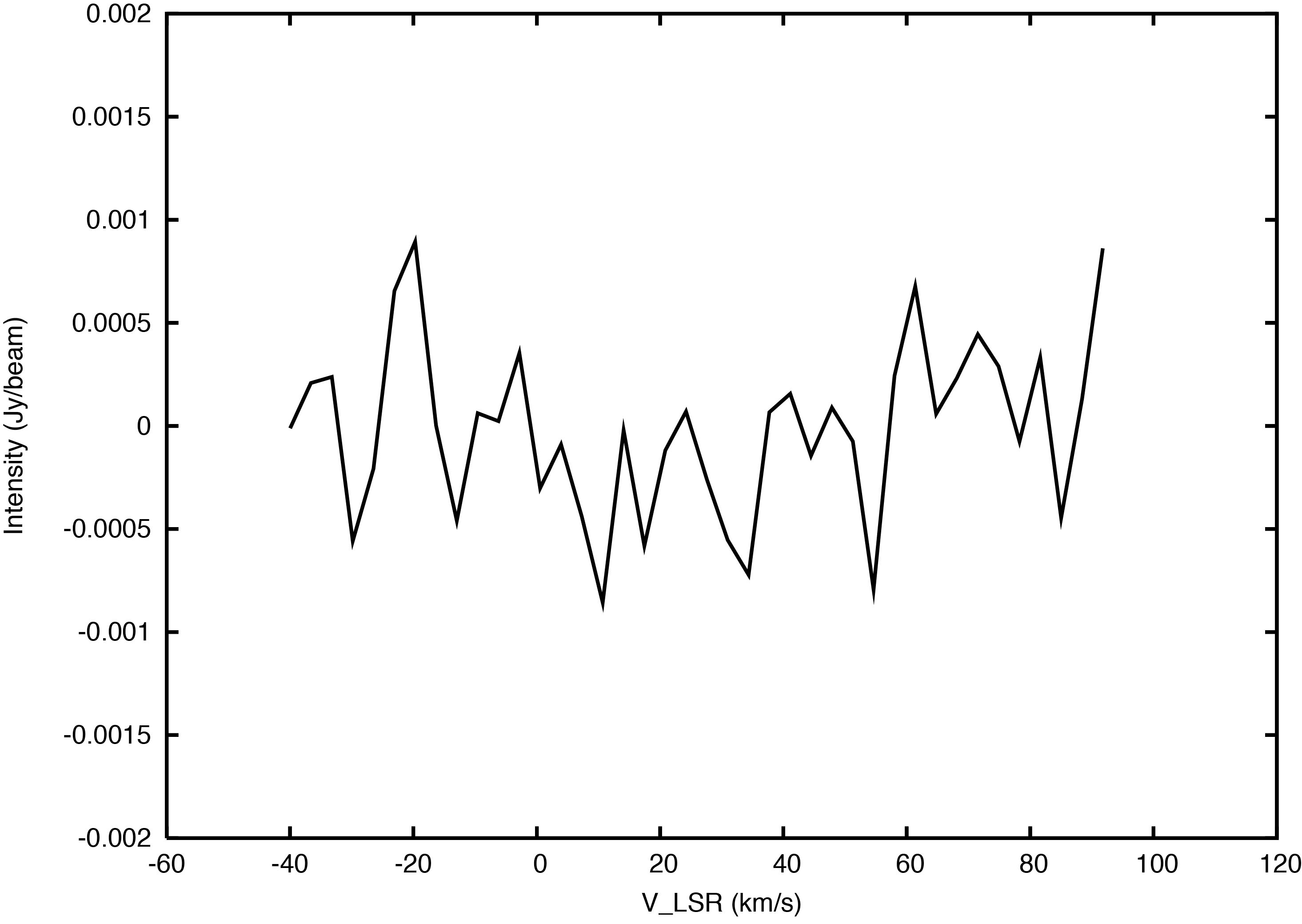}}
\vspace{0.5cm}
\hbox{\includegraphics[width=8.5cm]{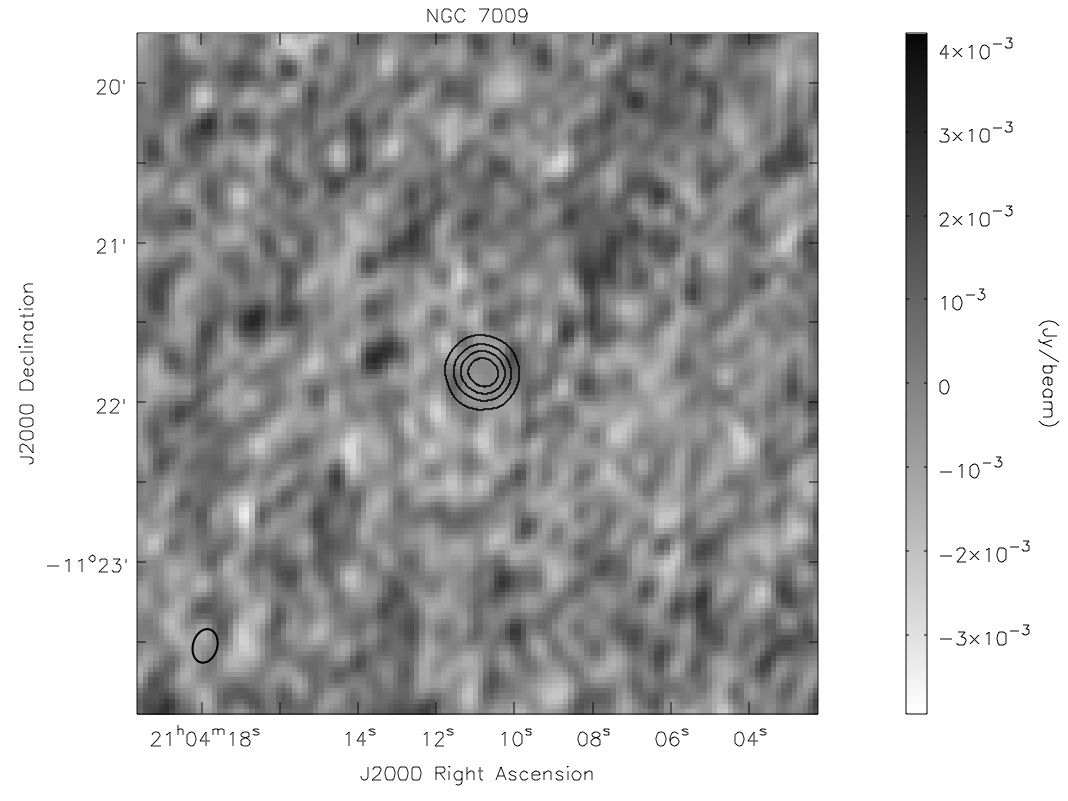}
\hspace{0.1cm}
\includegraphics[width=9cm]{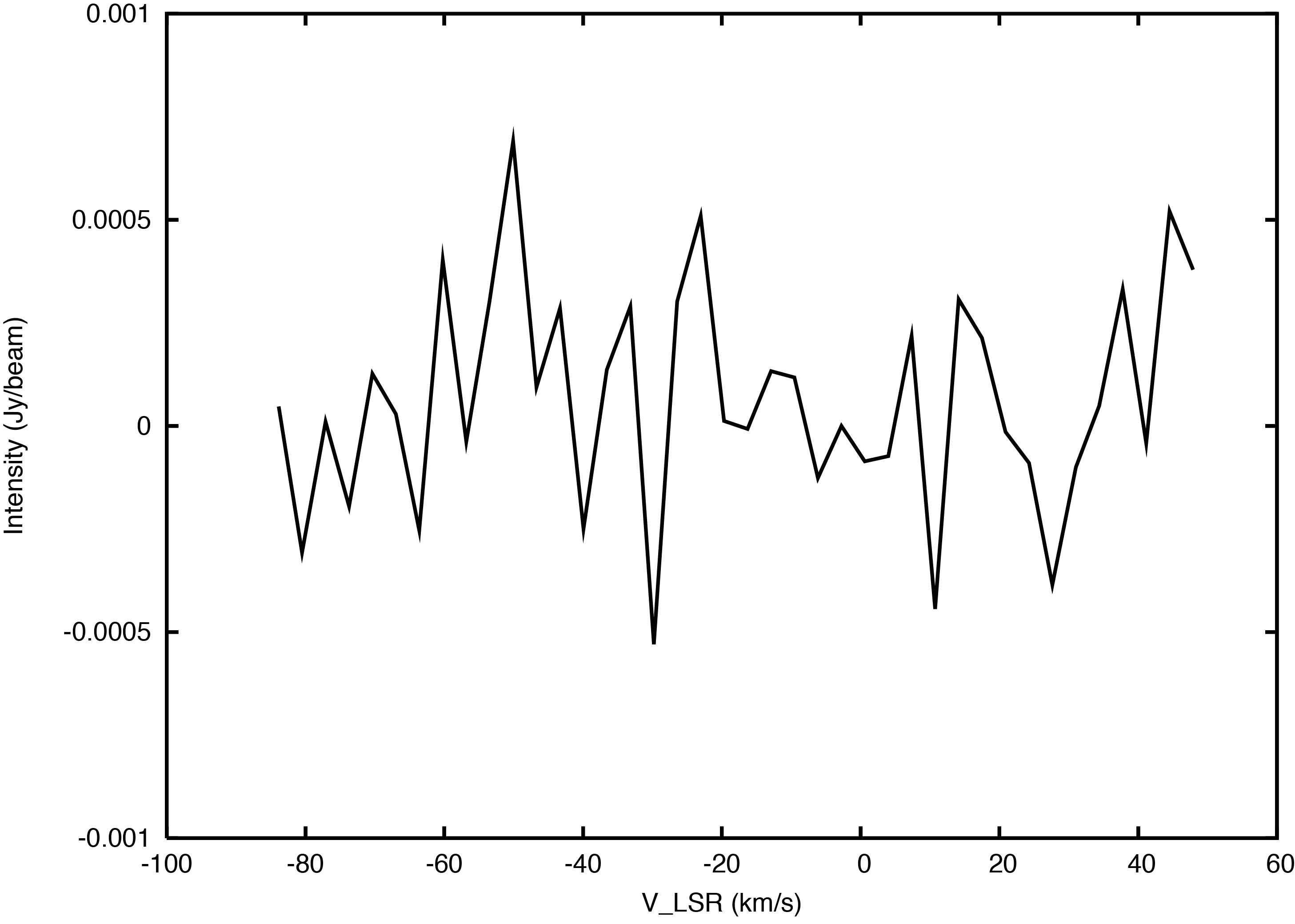}}
\caption[Line images]{{\it VLA} observations of IC 418 (top), NGC 6572 (middle), and NGC 7009 (bottom). On the left side the image shows the line emission of the source, continuum subtracted. The red contours in the centre show the  position of the continuum emission for each source. On the right side is the spectrum at the peak continuum position of the source. The $^3$He$^+$ line should be at 62 km\,s$^{-1}$ for IC 418, at -8.7 km\,s$^{-1}$ for NGC 6572 and at -46.6 km\,s$^{-1}$ for NGC 7009.}
\label{casa}
\end{figure*}

\section{Results}
The result of the observations are presented in Figure \ref{casa}, showing the emission cube with the continuum subtracted. As a reference the emission of the continuum is over plotted in red contours. The spectrum presented on the right of every image is the emission coming from the centre of the contours, where the emission of the line should be. The $^3$He$^+$ line was not detected for any of the objects, therefore only upper limits for the density values were calculated. We used a 3$\sigma$ level of the rms as an upper limit for the three sources.

The $^3$He$^+$ column density can be obtained using:
\begin{equation} 
N(^3\rm{He}^+) =\frac{\it{g_l} +\it{g_u}}{\it{g_u}}\frac{8\pi \it{k}\nu^2}{\it{hc^3\rm{A}_{\it{ul}}}}\int{T_B(\it{v})d\it{v}},
\end{equation}

\noindent where $g_u=$1 is the {\it g} value or magnetic moment of the upper state, $g_l=$3 is the {\it g} value or magnetic moment of the lower state, $A_{ul}=$1.95436$\times$10$^{-12}$s$^{-1}$ \citep{gould94}, and $T_B(v)$ is the brightness temperature profile of the line. For a gaussian line profile,

\begin{equation} 
\int{T_B(v)dv}=1/2\sqrt{\frac{\pi}{\rm{ln}2}}T_{B}^{0}\Delta{v},
\end{equation}

\noindent where $T_{B}^{0}$ is the brightness temperature at the centre of the line and $\Delta{v}$ is the full width at half power. The brightness temperature $T_{B}^{0}$ is related to the observed beam-averaged brightness temperature $T_L$, by \citep{rood95}

\begin{equation} 
T_{B}^{0}=T_L\frac{\theta_{b}^{2}+\theta_{s}^{2}}{\theta_{s}^{2}},
\end{equation}

\noindent where $\theta_b$ and $\theta_s$ are the beam and source angular radii.

The number of $^3$He$^+$ atoms per unit volume, $n$($^3$He$^+$), can be obtained by dividing the column density $N$($^3$He$^+$) by the averaged optical path, $<\Delta{s}>$, through the source. Representing a PN as a homogeneous sphere of radius R=$\theta_s$D, where D is the distance, the optical path at a position angle $\theta$ is   $\Delta{s}(\theta)=2\sqrt{R^2-(\theta D)^2}$, and the optical path averaged on the source results $<\Delta{s}>=\pi R/2 = \pi \theta_sD/2$.

The expression for  $n$($^3$He$^+$) is then
\begin{equation} 
n(^3\rm{He}^+) = 8\sqrt{\frac{\pi}{ln2}}\frac{\it{g_l} + \it{g_u}}{\it{g_u}}\frac{\it{k}\nu^2}{\it{hc^3}A_{\it{ul}}}\frac{\theta_{b}^{2}+\theta_{s}^{2}}{D\theta_{s}^{3}}T_L\Delta{\it{v}}.
\label{dens}
\end{equation}

Rearranging Equation \ref{dens} and using the numerical values, the final expression can be presented as follows,

\begin{eqnarray} 
\label{density}
n(^3\rm{He}^+) & = & 22.7 \left(\frac{T_L}{\rm{mK}}\right)\left(\frac{\Delta{\it{v}}}{\rm{km\,s^{-1}}}\right)\left(\frac{D}{\rm{kpc}}\right)^{-1} \\
\nonumber
& & \left(\frac{\theta_{s}}{{\rm arcsec}}\right)^{-3}\left(\frac{\theta_{b}}{41\arcsec}\right)^2\left(1+\frac{\theta_{s}^{2}}{\theta_{b}^{2}}\right)\rm{cm^{-3}}
\end{eqnarray}

Table \ref{parameters} shows the calculated $^3$He$^+$ density using Equation \ref{density} for the three sources. The distance values used were 1.2kpc for NGC 7009 \citep{galli97},  1.2kpc for NGC 6572 \citep{hajian95} and 1.3kpc for IC 418 \citep{me09}. To calculate the abundance we need to divide by the H$^+$ density. The H$^+$ values presented in Table \ref{parameters} were taken from \citet{galli97,balser06,morisset09} for NGC 7009, NGC 6572 and IC 418 respectively. The H$^+$ abundance is derived from the thermal free-free bremsstrahlung radio continuum emission, and is proportional to the square of the density. To obtain these values, the PNe where modelled using a numerical code, called NEBULA for the case of NGC 6572 and NGC 7009. This code calculates the radio recombination line and continuum emission from an ionised plasma. The modelled nebulae are composed solely of hydrogen and helium. The numerical grid uses a three-dimensional Cartesian coordinate system that allows for the specification of a completely arbitrary density and ionisation structure. For the case of IC 418 its H$^+$ abundance was modelled using Cloudy\_3D \citep{morisset06}.

\begin{table*}
\caption[PNe abundances]{Parameters used to calculate the $^3$He abundance in the PNe.}
\label{parameters}
\centering
\begin{tabular}{lccccccc} 
\hline\hline PNe & D & M &  T$_L$ & $\Delta v$ &  $n$(H$^+$) &  $n$($^3$He$^+$) & $^3$He$^+$/H$^+$ \\
     			  & (kpc) & (M$_{\odot}$) & (mK) & (km\,s$^{-1}$) & (cm$^{-3}$) & (cm$^{-3}$) & (10$^{-3}$)\\
\hline
IC 418	 & 1.3   & 1.7$\pm$0.3$^1$ & 27     &  30  & 3$\times$10$^3$    & $<$12.99   & $<$4.33\\	
NGC 6572 & 1.2  & 1.0$\pm$0.2$^2$ &150     &  15  & 22$\times$10$^3$     & $<$5.38     & $<$0.24 \\
NGC 7009 & 1.2  & 1.4$\pm$0.2$^3$ & 465    &  43  & 3.62$\times$10$^3$  & $<$29.85  & $<$8.24 \\		
 \hline
   \end{tabular}\\
$^1$\citet{morisset09}, $^2$\citet{hyung94}, $^3$\citet{galli97}   
\end{table*}

Once we calculated the $^3$He$^+$/H$^+$ abundance, we can compare the values with the theoretical calculations and the stellar models predictions.

Figure \ref{abundances} shows the $^3$He abundance of the three PNe from this work, they are presented using the red symbols. The mass values used are 1.7$\pm$0.3 for IC 418 \citep{morisset09}, 1.0$\pm$0.2 for NGC 6572 \citep{hyung94} and 1.4$\pm$0.2 for NGC 7009 \citep{galli97}.  
For comparison the PNe from \citet{balser97} are also presented (using green symbols), where the mass estimates are from \citet{galli97}. All these measurements represent upper limits points. For all the $^3$He$^+$/H$^+$ values calculated and taken from the literature it has been assumed that $^3$He/H$\simeq$ $^3$He$^+$/H$^+$ \citep{balser94}.

PNe IC 418 and NGC 7009 have values consistent with Population I and II theoretical predictions. For IC 418, this is the first upper limit calculation and it agrees with the stellar model predictions. For NGC 7009 a $^3$He/H abundance of $<8.24\times10^{-3}$ was calculated. Previous observations of this source gave estimates of $<(5.8-31)\times10^{-4}$, therefore our estimated value does not improve earlier estimates. In the case of NGC 6572 the abundance calculated is lower than the theoretical models predict. An upper limit for this PNe was calculated by \citet{balser06}: the value they estimate is $<10^{-3}$. We found an upper limit of $<2.4\times10^{-4}$.

\section{Discussion}
\begin{figure}
\centering
\includegraphics[width=8.5cm, height=7cm]{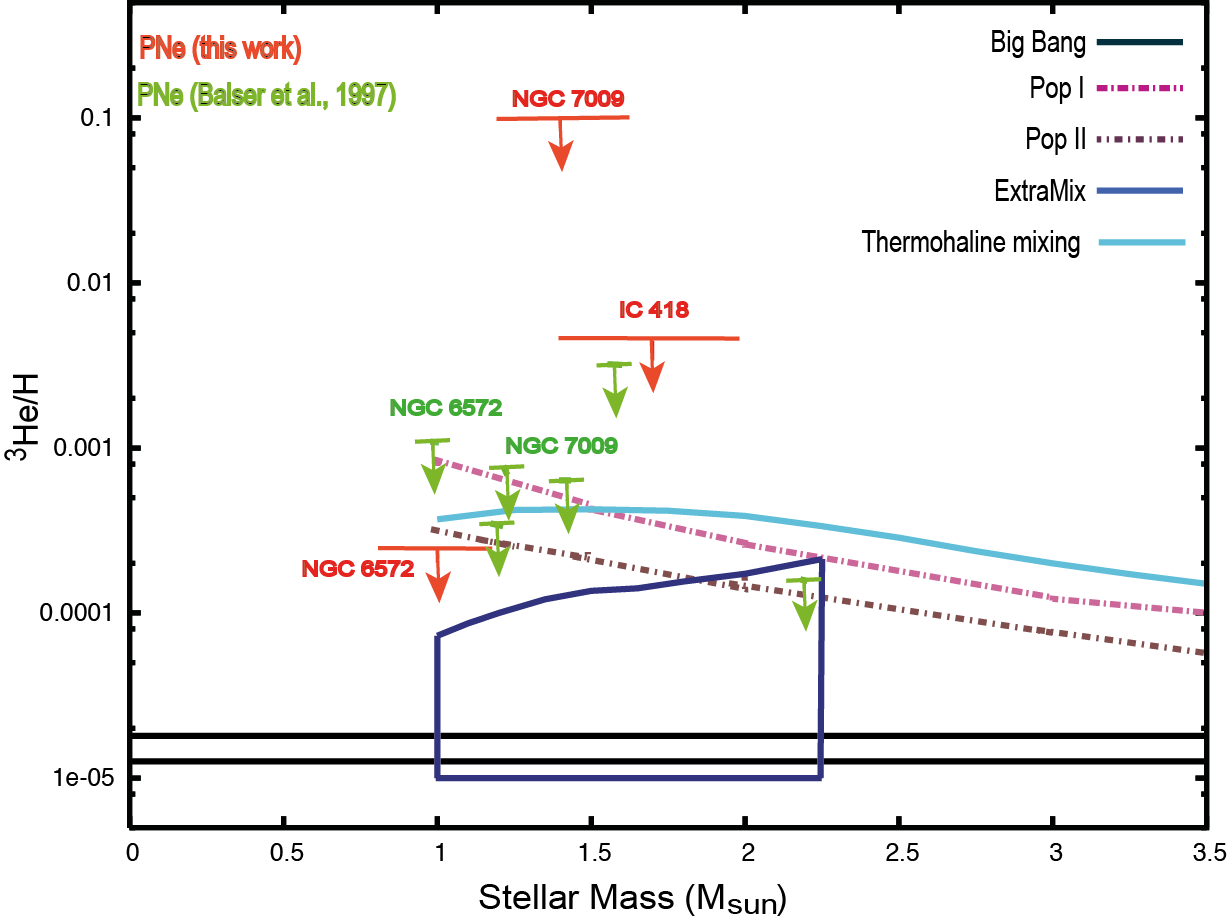}
\caption[$^3$He/H abundance]{Abundance of $^3$He/H versus main-sequence masses. The red symbols represent the work from this paper. The horizontal bar represents the mass range of the progenitor star. 
The green symbols represent the sample of 6 PNe (NGC 3242, NGC 6543, NGC 6720, NGC 7009, NGC 7662 and IC 289) from \citet{balser97}, using the mass estimates from \citet{galli97}. The curves as Pop I and II show the standard abundance of $^3$He taken from \citet{weiss96}. The blue curve labelled ExtraMix represents the results of stellar nucleosynthesis calculations using deep mixing by \citet{boothroyd99}. The light-blue curve represent the stellar models using thermohaline mixing. The Big Bang line is the primordial value of $^3$He/H.}
\label{abundances}
\end{figure}

The stellar evolution models are also presented in Figure \ref{abundances}. The curves labelled as Pop I and II show the standard abundance of $^3$He taken from \citet{weiss96}. The Big Bang line is the primordial value of $^3$He/H. 

In order to understand the chemical evolution of the Galaxy, it is crucial we understand the contribution of PNe to the $^3$He abundance. Chemical evolution models started including only processed primordial deuterium (D) as a source of $^3$He synthesis in low mass stars. But even without including any stellar production of $^3$He, these models showed that $^3$He/H increases with time \citep{vangioni94}.
Because the abundance of $^3$He in H\,{\sc ii} regions, the ISM and the proto-solar system is much lower than predicted, new chemical models were proposed. 

These models include non convective mixing processes that either prevent the buildup of $^3$He on the main-sequence or destroys $^3$He along the upper red giant branch \citep{boothroyd99,lagarde12}. 

The blue curve labelled ExtraMix represents the results of stellar nucleosynthesis calculations using deep mixing by \citet{boothroyd99}. 
The detailed calculations indicate that the mass dependence of the destruction of $^3$He is very strong, and the resulting abundance decreases sharply for masses below $\sim$2.5M$_{\odot}$. It is clearly seen in this Figure that the observed PNe have not suffered any depletion. Almost all of the PNe are more than a factor of $\sim$2 above the non-standard curve (the Extramix predictions). 

One suggestion to solve this is by adding some additional mixing
mechanism during the first ascent of the RGB. By
allowing material to circulate below the convective envelope after the first
dredge-up has occurred, the surface abundances can be modified as the
star approaches the tip of the RGB. Observationally, the existence of
this non-convective mixing is well established (e.g. \citealp{gratton00}) as carbon, lithium and the $^{12}$C/$^{13}$C ratio are all
observed to change as stars reach the tip of the RGB. That this process
should also affect the $^3$He abundance seems clear as to affect these
elements and isotopes one must reach temperatures where $^3$He burning
reactions will occur. However, the physical cause of this non-convective
mixing has proven elusive and many possible mechanisms, such as rotation
(e.g. \citealp{palacios06}) have been suggested.

The hydrodynamic simulations of \citet{eggleton06} pointed to a
particularly convenient process for this extra mixing. Their simulations
showed that a mean molecular weight inversion caused by the destruction
of $^3$He could drive circulation of material. This process
(subsequently identified with thermohaline convection by \citealp{charbonnel07}) would allow low-mass stars to destroy the $^3$He that they produce during the main sequence phase.

Newer stellar models from \cite{lagarde12} analyse the evolution of $^3$He and its destruction due to the effects of thermohaline instability and rotation-induced mixing. Their models show that thermohaline mixing can induce significant depletion of $^3$He in low- and intermediate-mass stars. Although these stars remain the net producers of $^3$He, their contribution to the Galactic evolution of this element is highly reduced compared to classical theory. It is claimed in their paper that thermohaline mixing is the only physical mechanism known so far able to solve the so-called ``The $^3$He Problem". Importantly, its inhibition by a fossil magnetic field in red giant stars that are descendants of Ap stars does reconcile the measurements of $^3$He/H in Galactic H\,{\sc ii} regions with high values of $^3$He in a few PNe. 

We have used the STARS stellar evolution code \citep{eggleton71,pols95,stancliffe09} to produce a set of stellar models in the mass range 1-3.5 solar masses at intervals of 0.25 solar masses and with a metallicity of Z=0.02. These models have been computed from the pre-main sequence to the first thermal pulse on the asymptotic giant branch. Thermohaline mixing has been included at all stages of the evolution using the diffusive formalism of \citet{ulrich72} and \citet{kippen80}. Following the work of \citet{charbonnel07}, we set the free parameter of the formalism to $C_t = 1000$, a value which not only reproduces the compositional changes in field giants, but which has also been shown to reproduce the difference in behaviour between C-normal and C-rich metal-poor stars \citep{stancliffeB09} as well as the abundance changes in globular cluster stars \citep{angelou11}. Note that this value is not supported by current hydrodynamical simulations of thermohaline mixing \citep{traxler11, deni011} and that some other process (or processes) could be the cause of abundance changes on the giant branch (e.g. magnetic fields, \citealp{charbonnel07b, nordhaus08}). These stellar models using thermohaline mixing are also represented in Figure \ref{abundances} (light-blue line). 

As shown in Figure \ref{abundances} the $^3$He abundance of two PNe (NGC 6572 and NGC 6720) fall below the line that the model predicts, therefore at least for these two PNe the thermohaline mixing model fails to reproduce their $^3$He abundances.

\section{Conclusions}

Upper-limits for the $^3$He abundance of three PNe have been estimated using the $^3$He$^+$ emission line observed with the {\it VLA} towards the PNe IC 418, NGC 6572 and NGC 7009. In total one strong detection (PN J320) and seven upper limits of the $^3$He abundance on PNe have been reported in the literature, including this work, and are consistent with standard stellar evolution calculations. These estimates are two orders of magnitude larger than what is found in H\,{\sc ii} regions \citep{balser97}, the local ISM \citep{gloecker96} and the proto-solar system \citep{geiss93}. 

The resulting evolution of $^3$He is only consistent with the values determined in pre-solar material and the ISM if more than 96\% of the whole population of stars with mass below $\sim$2.5M$_{\odot}$ have undergone enhanced $^3$He depletion. These results suggest that more observations are needed in order to make a strong conclusion about the stellar evolution models. 

\section*{Acknowledgments}
This work is based on observations made with the {\it VLA} from NRAO, PI A. Zijlstra. LGR and AZ acknowledge the support of CONACyT (Mexico). Astrophysics at Manchester is supported by  grants from the STFC. JEP has received funding from the European CommunityÕs Seventh Framework Programme (/FP7/2007-2013/) under grant agreement No 229517. RJS is the recipient of a Sofja Kovalevskaja Award from the Alexander von Humboldt Foundation. 

\newcommand{\mockalph}[1]{}

\bibliographystyle{mn2e}
\addcontentsline{toc}{chapter}{\bibname}
\bibliography{biblioThesis}
\bsp

\label{lastpage}

\end{document}